\documentclass[12pt]{article}

\usepackage{latexsym}
\usepackage{amssymb,amsfonts,amsmath}
\usepackage{graphicx} 
\usepackage{indentfirst}
\usepackage{bbm}
\usepackage{amssymb}
\usepackage{verbatim}
\usepackage{amsmath, amsthm,amssymb}
\usepackage{mathrsfs}
\usepackage{slashed}
\usepackage{hyperref}
\usepackage{amsfonts}
\usepackage{dsfont}
\usepackage{enumerate}

\parskip=\medskipamount

\newcommand {\cN}{{\cal N}}

\def\a{\alpha}

\def\b{\beta}

\def\d{\delta}

\def\f{\phi}
\def\g{\gamma}

\def\j{\psi}

\def\l{\lambda}
\def\m{\mu}
\def\n{\nu}

\def\s{\sigma}

\def\x{\xi}
\def\z{\zeta}

\def\F{\Phi}

\def\P{\Pi}

\def\rd{{\rm d}}
\def\ri{{\rm i}}

\newcommand{\ad}{{\dot{\alpha}}}                           
\newcommand{\bd}{{\dot{\beta}}}                            
\newcommand{\ve}{\varepsilon}                            

\newcommand{\ab}{{\a\b}}

\newcommand{\pa}{\partial}                           
\newcommand{\hf}{\frac12}

\newcommand{\vf}{\varphi}

\newcommand{\be}{\begin{equation}}
\newcommand{\ee}{\end{equation}}
\newcommand{\bea}{\begin{eqnarray}}
\newcommand{\eea}{\end{eqnarray}}
\newcommand{\non}{\nonumber}
\newcommand{\bm}[1]{\mbox{\boldmath$#1$}}

\def\double #1{#1{\hbox{\kern-2pt $#1$}}}

\newcommand{\sSU}{\mathsf{SU}}
\newcommand{\sSL}{\mathsf{SL}}
\newcommand{\sGL}{\mathsf{GL}}
\newcommand{\sSO}{\mathsf{SO}}

\newcommand{\sU}{\mathsf{U}}

\newif\ifdtup

\newcommand{\bsubeq}{\begin{subequations}}
\newcommand{\esubeq}{\end{subequations}}

\numberwithin{equation}{section}

\begin{document}

\begin{titlepage}
\begin{flushright}
December, 2018\\
\end{flushright}
\vspace{5mm}

\begin{center}
{\Large \bf Spin projection operators and  higher-spin Cotton tensors
in three dimensions}\\ 
\end{center}

\begin{center}

{\bf 
Evgeny I. Buchbinder,  
Sergei M. Kuzenko, James La Fontaine\\
 and Michael Ponds 
} \\
\vspace{5mm}

\footnotesize{
{\it Department of Physics M013, The University of Western Australia\\
35 Stirling Highway, Crawley W.A. 6009, Australia}}  
\vspace{2mm}
~\\
Email: \texttt{evgeny.buchbinder@uwa.edu.au, 
sergei.kuzenko@uwa.edu.au, 21319182@student.uwa.edu.au, michael.ponds@research.uwa.edu.au}\\
\vspace{2mm}

\end{center}

\begin{abstract}
\baselineskip=14pt
We elaborate on the spin projection operators in three dimensions 
and use them to derive a new representation for the linearised higher-spin Cotton tensors. 

\end{abstract}

\vfill

\vfill
\end{titlepage}

\newpage
\renewcommand{\thefootnote}{\arabic{footnote}}
\setcounter{footnote}{0}

\tableofcontents{}
\vspace{1cm}
\bigskip\hrule

\allowdisplaybreaks


\section{Introduction} 

In four dimensions (4D), there exists a remarkably simple expression for the linearised 
higher-spin Weyl tensors in terms of gauge prepotentials 
$h_{\a_{1} \dots \a_{s}  \ad_1 \dots \ad_s} = h_{(\a_{1} \dots \a_{s} )( \ad_1 \dots \ad_s)}$, 
see e.g. \cite{FL1,FL2} and section 6.9 of \cite{BK}.
In the case of an integer spin $s \geq 1$, 
it reads
\bea
{C}_{\a_1 \dots \a_{2s}} = \pa_{(\a_1}{}^{\bd_1} \dots \pa_{\a_s}{}^{\bd_s}
h_{\a_{s+1} \dots \a_{2s} ) \bd_1 \dots \bd_s}~.
\label{0.1}
\eea
For a half-integer spin $s+\hf \geq \frac 32$, with $ s = 1,2, \dots,$
we  have
\bea
{C}_{\a_1 \dots \a_{2s+1}} = \pa_{(\a_1}{}^{\bd_1} \dots \pa_{\a_s}{}^{\bd_s}
\j_{\a_{s+1} \dots \a_{2s+1} ) \bd_1 \dots \bd_s}~.
\label{0.2}
\eea
It follows from \eqref{0.1} and \eqref{0.2} that the higher-spin Weyl tensors
are invariant under gauge transformations of the form 
\bea 
\d_\z h_{\a_{1} \dots \a_{s}  \ad_1 \dots \ad_s} &=&\pa_{(\a_1 (\ad_1} 
\z_{\a_2 \dots \a_s) \ad_2 \dots \ad_s) } ~ , \label{0.03}\\
\d_\x \j_{\a_{1} \dots \a_{s+1}  \ad_1 \dots \ad_s} &=&\pa_{(\a_1 (\ad_1} 
\x_{\a_2 \dots \a_{s+1}) \ad_2 \dots \ad_s) } ~.
\eea
It should be remarked that there are two ways for the bosonic gauge field
$h_{\a(s) \ad(s)} = h_{\a_{1} \dots \a_{s}  \ad_1 \dots \ad_s}$
to occur in higher-spin gauge theories.\footnote{The story with the fermionic gauge field
$\j_{\a(s+1) \ad(s)}$ is analogous.}
Firstly, $h_{\a(s) \ad(s)} $ is one of the two gauge 
prepotentials $\{ h_{\a(s) \ad(s)} ,h_{\a(s-2) \ad(s-2)} \}$ 
in the Fronsdal massless integer-spin models\footnote{The compensator $h_{\a(s-2) \ad(s-2)}$
transforms under \eqref{0.03} by the rule $\d_\z h_{\a(s-2) \ad(s-2)} \propto
\pa^{\b \bd} \z_{\b \a (s-2) \bd \ad(s-2)}$.}
 \cite{Fronsdal1,Fronsdal2}
(see section 6.9 of \cite{BK} for a review). Secondly, it is the  gauge field
in the Fradkin-Tseytlin conformal integer-spin theories \cite{FT}.
In the former theories, $C_{\a(2s)} $ and its conjugate are the only gauge invariant field
strengths which survive on the mass shell. In the latter theories, the gauge-invariant 
action may be formulated in terms of $C_{\a(2s)} $ and its conjugate
\cite{FL1,FL2}.

In three dimensions,
the Weyl tensor vanishes identically
and all information about the conformal geometry 
of spacetime is encoded in 
 the Cotton tensor. Spacetime is conformally flat if and only if the Cotton tensor vanishes
\cite{Eisen}
(see \cite{BKNT-M1} for a modern proof). Linearised higher-spin extensions
of the Cotton tensor in Minkowski space were constructed in \cite{PopeTownsend}
and  \cite{K16}
in the bosonic and fermionic cases, respectively.
In terms of a gauge prepotential $h_{\a_1 \dots \a_n} =h_{(\a_1 \dots \a_n)}$, with $n>1$,
 the linearised Cotton tensor is given by the expression \cite{K16}
 \begin{align} 
 C_{\a(n)}(h)=\frac{1}{2^{n-1}}\sum_{j=0}^{\lfloor n/2\rfloor}\binom{n}{2j+1}\Box^j\partial_{(\a_1}{}^{\b_1}\dots\partial_{\a_{n-2j-1}}{}^{\b_{n-2j-1}}
 h_{\a_{n-2j}\dots\a_n)\b_1\dots\b_{n-2j-1}}~,
 \label{0.3}
 \end{align} 
where $\lfloor x\rfloor$ denotes the floor function and returns the integer part of a real number $x\geq 0$. The fundamental properties of $C_{\a(n)}$ are the following: 
\begin{enumerate}[(i)]
\item $C_{\a(n)}$
is invariant under gauge 
transformations of the form
\bea
\d_\z h_{\a(n)} = \pa_{(\a_1\a_2} \z_{\a_3 \dots \a_n)} \quad \implies \quad \d_\z C_{\a(n)}=0~;
\label{0.4}
\eea
\item  $C_{\a(n)}$ is divergenceless 
\bea
\pa^{\b\g} C_{\b\g \a_1 \dots \a_{n-2}} =0~.
\label{0.5}
\eea
\end{enumerate}

Unlike the 4D relations \eqref{0.1} and \eqref{0.2}, the expression for $C_{\a(n)}$ given by 
\eqref{0.3} is not illuminating. It is not obvious from \eqref{0.3} that $C_{\a(n)}$ possesses 
the properties \eqref{0.4} and \eqref{0.5}.
Recently it has been shown, first  in the bosonic (even $n$)  \cite{HHL} and later in the fermionic
 (odd $n$) \cite{HLLMP} case, that \eqref{0.3}
is the most general solution of the conservation equation \eqref{0.5}, and the proofs are non-trivial.
There exists a simple proof of this statement based on the use of $\cN=1$ 
supersymmetry \cite{K16}. 
However, it makes use of an embedding of the higher-spin gauge 
prepotentials in superfields. A simple non-supersymmetric proof of this statement  is still missing. 

In this letter we derive a new representation for the higher-spin Cotton tensor $C_{\a(n)}$
which  is analogous to the 4D relations \eqref{0.1} and \eqref{0.2} and which makes 
obvious the properties \eqref{0.4} and \eqref{0.5}. Our approach is based on the use 
of 3D analogues of the  Behrends-Fronsdal projection operators  \cite{BF,Fronsdal}
(see \cite{SG,GGRS,Isaev:2017nud} for modern descriptions using the two-component
spinor formalism).
These projection operators were generalised beyond four dimensions by Segal
\cite{Segal} (for recent discussions, see also \cite{Isaev:2017nud,Bonezzi}) for integer spin values, while the half-integer-spin case was
described in \cite{Isaev:2017nud}. 
As will be shown below, the 3D case is somewhat special.

This paper is organised as follows. In section 2 we discuss various aspects of massive higher-spin fields. Section 3 is devoted to spin projection operators. 
In section 4 we derive a new representation for the higher-spin Cotton tensor 
 $C_{\a(n)}$.  Concluding comments are given in section 5. Our spinor conventions 
 are summarised in the appendix. 


\section{On-shell massive  fields in three dimensions }


We start with discussing tensor fields realising 
irreducible massive (half-)integer spin
representations of the Poincar\'e group in three dimensions.
We restrict our attention to the case of integer and half-integer spin values;
for a discussion of the anyon representations see, e.g., \cite{JN}.
The 3D spin group\footnote{${\sSL} (2, {\mathbb R})$ is 
a double covering of the connected Lorentz group $\sSO_0(2,1)$.
The universal covering group of ${\sSL} (2, {\mathbb R})$ is not a matrix group 
and cannot be embedded in any group
$\sGL (n, {\mathbb R})$, see \cite{NT} for the proof.}
is ${\sSL} (2, {\mathbb R})$, so that the fields of interest are real symmetric rank-$n$
 spinors, 
 $\Phi_{\a_1 \dots \a_n} =\F_{(\a_1 \dots \a_n)}
 \equiv \Phi_{\a (n)}$.

For $n>1$, an on-shell  field $\Phi_{\a (n)}(x) $ of mass $m$
 satisfies the following differential equations 
 \cite{GKL,TV} (see also \cite{BHT}):
\begin{subequations}\label{1.1.12}
\bea
\pa^{\b \g} \Phi_{\b \g \a (n-2)}&=&0\,, 
\label{1.1.1}
\\
\pa^\b{} _{(\a_1} 
\Phi_{\a_2 \dots \a_n) \b} &=& m \s \Phi_{\a (n)}\,, \qquad \s =\pm 1 \,. 
\label{1.1.2}
\eea
\end{subequations}
In the spinor case, $n=1$, eq.  \eqref{1.1.1} is absent, and it is the Dirac equation 
\eqref{1.1.2} which defines a massive field. 
The constraints \eqref{1.1.1} and  \eqref{1.1.2}
imply the mass-shell equation\bea
(\Box -m^2 ) \F_{\a (n)} =0~.
\label{mass-shell}
\eea
Equations \eqref{1.1.1} and 
\eqref{mass-shell} prove to be equivalent to the 3D Fierz-Pauli field equations \cite{FP}.
It is worth pointing out that the equations \eqref{1.1.12} naturally 
originate upon quantisation of the  particle models studied 
in \cite{GKL,deAzcarraga:2014hda}. 

Let $P_a$ and $J_{ab}= -J_{ba}$ be the generators of the 3D Poincar\'e group.
The Pauli-Lubanski pseudo-scalar 
\bea
W:= \hf \ve^{abc}P_a J_{bc} = -\hf P^{\a\b} J_{\a\b}
\label{PauliL}
\eea
commutes with the generators $P_a$ and $J_{ab}$.
Irreducible unitary representations of the Poincar\'e group 
are labelled by two parameters,  mass $m$ and helicity $\l$, 
which are associated with the Casimir operators, 
\bea
 P^a P_a = -m^2 {\mathbbm 1} ~, \qquad W=m \l {\mathbbm 1}~.
 \label{Casimirs}
 \eea
 The parameter $|\l|$ is identified with spin. 
 
In the case of field representations, we have
\bea
W= \hf \pa^{\ab} M_{\a\b}~,
\eea
where the action of $M_{\a\b}=M_{\b\a}$ on a field 
$\F_{\g(n)}$ is defined by 
\bea
M_{\a\b} \F_{\g_1 \cdots \g_n} = \sum_{i=1}^n
\ve_{\g_i (\a} \F_{\b) \g_1 \cdots \widehat{\g_i} \dots\g_n}~,
\eea
where the hatted index of $\F_{\b \g_1 \cdots \widehat{\g_i} \dots\g_n}$  is omitted.
It follows from \eqref{1.1.2} and the second relation in \eqref{Casimirs} that 
the helicity of the on-shell massive field $\F_{\a(n)}$ is 
\bea
\l = \frac{n}{2} \s~.
\eea

In order to make contact with Wigner's classification of unitary representations of the Poincar\'e
group \cite{Wigner} and its 3D extension \cite{Binegar}, 
it is more convenient to work in momentum space in which 
 the equations \eqref{1.1.12} take the form
\begin{subequations}\label{1.2.12}
\bea
&&
p^{\b \g} \Phi_{\b \g \a (n-2)}(p) =0\,, 
\label{1.2.1}
\\
&&
p^\b{}_{(\a_1} 
\Phi_{\a_2 \dots \a_n )\b} (p) = - {\rm i} \s m  \Phi_{\a (n)} (p) \,, \qquad \s =\pm 1 \,,
\label{1.2.2}
\eea
\end{subequations}
where $\Phi_{\a (n)} (p)$ denotes the positive-energy part of the Fourier transform of 
 $\Phi_{\a (n)} (x)$. 
We now develop some group-theoretical aspects  before discussing the  equations
\eqref{1.2.12} in  more detail.

Let $q^a = (m, 0, 0)$ be the momentum of a massive particle at rest. Then an arbitrary  three-momentum $p^a$ of the  particle is obtained by applying a proper orthochronous Lorentz transformation 
  to $p^a$, that is 
\be 
(p\cdot \g)_{\a\b}:=
p^a (\gamma_a)_{\a \b} \equiv  p_{\a \b} = (L (q \cdot \gamma) L^T)_{\a \b}\,, 
\label{1.8}
\ee
for some matrix 
$L\in  \sSL(2, {\mathbb R})$. It is convenient to parametrise $L$ in terms of two
linearly independent real commuting spinors
\be 
L=\frac{1}{(\nu , \mu)^{1/2}}
 \left(
\begin{array}{cc}
 \mu_1 & \nu_1  \\
 \mu_2 & \nu_2 
\end{array}
\right) = \frac{1}{(\nu , \mu)^{1/2}} (\mu_{\a}, \nu_{\a})\,, \qquad 
(\nu , \mu) :=\nu^{\a} \mu_{\a} = -(\mu , \nu)\, .
\label{1.9}
\ee
Here the spinors 
 $\mu_{\a}$ and $\nu_{\a}$ are arbitrary modulo the condition
 $(\nu , \mu)  >0$. 
Note that $\det  L=1$. 
It should be remarked that \eqref{1.9} is  
invariant under the rescalings $\mu_{\a} \to \rho \mu_{\a}, \ \nu_{\a} \to \rho \nu_{\a}$. 
In principle, we can use this symmetry 
to normalise $(\nu , \mu) =1$, but we prefer to keep all expressions in the most general form.

Making use of the relations \eqref{1.8} and \eqref{1.9} gives
\be
p_{\a \b}= \frac{m}{(\nu , \mu)} (\mu_{\a} \mu_{\b} + \nu_{\a} \nu_{\b})\,. 
\label{1.10}
\ee
The identities 
$
(\mu , \mu)= (\nu , \nu)=0
$
imply that 
\be 
\hf p^{\a \b} p_{\a \b}= - p^a p_a =  m^2\,. 
\label{1.11}
\ee
Since $(q \cdot \g)_{\a \b}= m (\gamma_0)_{\a \b}=m {\mathbbm 1} $, 
it follows that $p_{\a\b} $ given by \eqref{1.8}  is invariant under the 
transformation $L \to L \cdot h$, where $h \in \sSO(2)$. 
The latter group is the 3D little group in the massive case. 

Since the little group $\sSO(2)$ is abelian,  Wigner's wave function $\f^{(\l)}(p)$, which describes 
the irreducible massive representation of helicity $\l$, must be one-component.
It follows from \eqref{1.2.12} that $\F_{\a(n)} (p)$ describes one degree of freedom
(it suffices to consider the $p^a=q^a$ case). However, even for the simplest choice
$p^a=q^a$ all components of $\F_{\a(n)}$ are non-vanishing. 
It would be convenient to have an approach that provides a simple rule to read off
a one-component Wigner wave function for every (half-)integer helicity.
For this we will use the isomorphism 
between $\sSL(2, {\mathbb R})$ and $\sSU(1, 1)$
described in detail in  \cite{Vilenkin}. 
Associated with a group element  $L = (L_{\a}{}^{\b} )\in \sSL(2, {\mathbb R})$
is the matrix $\tilde{L} =(\tilde{L}_\a{}^\b)  \in \sSU(1, 1)$ given by\footnote{Strictly speaking, 
different types of indices have to be used for the elements 
of  $\sSL(2, {\mathbb R})$ and $\sSU(1, 1)$. To avoid a cluttered notation, 
we will not make such a distinction. We simply denote all operators and tensors 
of $\sSU(1, 1)$ with a tilde.}
\be 
\tilde{L}= T^{-1} L T\,, 
\label{1.3}
\ee
where $T$ denotes the following unitary, unimodular matrix
\be 
T=\frac{1}{\sqrt{2}}
 \left(
\begin{array}{cc}
 1 & {\rm i}  \\
 {\rm i} & 1 
\end{array}
\right) \in \sSU(2)
\,.
\label{1.4}
\ee
If $\psi_{\a}$ is a spinor of $\sSL(2, {\mathbb R})$, the corresponding spinor $\tilde{\psi}_{\a}$ of $\sSU(1, 1)$ is given by $\tilde{\psi} = T^{-1} \psi$.
More generally,  associated with an arbitrary symmetric rank-$n$  $\sSL(2, {\mathbb R})$ spinor
 $\Phi_{\a (n)}$ is 
the  $\sSU(1, 1)$ tensor $\tilde{\Phi}_{\a (n)}$ defined by 
\be 
\tilde{\Phi}_{\a_1 \dots  \a_n} = (T^{-1})_{\a_1}{}^{ \b_1} \dots (T^{-1})_{\a_n}{}^{ \b_n} \Phi_{\b_1 \dots  \b_n} \,.
\label{1.5}
\ee
Hence, in the $\sSU(1, 1)$ picture the dynamical equations \eqref{1.2.1} and \eqref{1.2.2} look the same except that $p_{\a\b} $ and  $\Phi_{\a(n)}$ are replaced with   $\tilde{p}_{\a\b} $ and 
$ \tilde{\Phi}_{\a(n)} $, respectively.

We will parametrise the group elements  $\tilde{L} \in\sSU(1, 1)$ 
in terms of two  complex spinors $\tilde{\mu}_\a$ and $\tilde{\n}_\a$
that are related to each other by Dirac conjugation. More specifically, 
every element of $\sSU(1,1)$ can be represented as 
\be 
\tilde{L}=
 \left(
\begin{array}{cc}
 \tilde{\mu}_1 & \tilde{\nu}_1  \\
 \tilde{\mu}_2 & \tilde{\nu}_2 
\end{array}
\right) = 
(\tilde{\mu}_{\a}, \tilde{\nu}_{\a})\,, 
\qquad \tilde{\n}^\a = \overline{\tilde \m}{}^\a~,  
\qquad \tilde{\nu}^\a \tilde{\mu}_\a=1 \, .
\label{1.12}
\ee
where the Dirac conjugate $\overline{\tilde \j} =( \overline{\tilde \j}{}^\a)$ of a spinor 
${\tilde \j}= ({\tilde \j}{}_\a)$ is defined by 
\bea
\overline{\tilde \j} = {\tilde \j}{}^\dagger \s_3~.
\eea
We can rewrite ${\tilde L} $ in the form
\bea
{\tilde L} = (\tilde{\mu}_{\a}{}^+ , \tilde{\mu}_{\a}{}^-) \in \sSU(1,1) \, ,
\eea 
where `plus' and `minus' refer to charges with respect to the $\sU(1) $ action
\bea
\tilde{L} \to \tilde{L} \exp \big( \ri \vf \s_3 \big)~, \qquad \vf \in {\mathbb R}~.
\eea
With this notation the $\sSU(1,1)$ formalism is analogous to the $\sSU(2)$ one used within 
the harmonic superspace approach in four dimensions \cite{GIKOS}.

In the $\sSU(1, 1)$ picture, the momentum $\tilde{p}_{\a\b} = (p \cdot \tilde \g )_{\a\b}$ 
is obtained from ${p}_{\a\b} = (p \cdot  \g )_{\a\b}$ by the rule 
\bea
\tilde{p}_{\a \b} &= & (T^{-1})_{\a}{}^{ \g}(T^{-1})_{\b}{}^{ \d} p_{\g \d}= (T^{-1} p T^{-1})_{\a \b} 
=(\tilde{L} \tilde{q} \tilde{L}^{{\rm T}})_{\a \b}\,,
\label{1.14}
\eea
where the momentum of a particle at rest,
$\tilde{q}_{\a\b} = (q \cdot \tilde \g )_{\a\b}$,  becomes
\be 
\tilde{q}= T^{-1} (q \cdot \gamma) T^{-1} = m (T^{-1} )^2 = - {\rm i} m \sigma_1\,. 
\label{1.15}
\ee
Making use of  eqs.~\eqref{1.12} and~\eqref{1.15} gives
\be 
\tilde{p}_{\a \b} = -{\rm i} m
\big(\tilde{\mu}_{\a} \tilde{\nu}_{\b} + \tilde{\mu}_{\b} \tilde{\nu}_{\a} \big)\,. 
\label{1.17}
\ee
The stability group of $\tilde q$ consists of all group elements ${\tilde h} \in \sSU(1,1)$
with the property
\be 
\tilde{q} = \tilde{h} \tilde{q} \tilde{h}^{{\rm T}} \qquad \Longleftrightarrow \qquad 
\s_1= \tilde{h}\s_1 \tilde{h}^{{\rm T}}\,. 
\label{1.16}
\ee
Hence, the little group in the $\sSU(1,1)$ picture consists of the matrices
\be 
\tilde{h} = {\rm e}^{{\rm i} {\varphi} \s_3} \in \sSU(1, 1)\,, \qquad 
{\varphi} \in {\mathbb R}\,, 
\label{1.16.1}
\ee
and is isomorphic to $\sU(1)$. 

The group element $\tilde{L} \in \sSU(1,1)$ in \eqref{1.14} is defined modulo 
arbitrary right shifts
\bea
\tilde{L} \to \tilde{L} {\rm e}^{{\rm i} {\varphi} \s_3}  \,, \qquad 
{\varphi} \in {\mathbb R}\,.
\label{2.25}
\eea
This freedom may be fixed by choosing the global coset representative
\bea 
\tilde{L}(p)  =\frac{1}{\sqrt{2m(p^0+m)}}
 \left(
\begin{array}{cc}
 p^0+m  ~& ~ p^1 +{\rm i} p^2  \\
p^1-{\rm i} p^2  ~&  ~  p^0+m
\end{array}
\right) \in \sSU(1,1)
\,,
\eea
which parametrises the homogeneous space $\sSU(1,1) / \sU(1)$  that is  diffeomorphic to
the hyperbolic plane ${\mathbb H}^2$.

Now we are prepared to construct massive fields $ \tilde{\Phi}^{(\pm)}_{\a (n)}$
of helicity $\pm n/2$. They are:
\begin{subequations}\label{1.18}
\bea
&& 
\tilde{\Phi}_{\a_1 \a_2 \dots \a_n}^{(+)} ({p}) = \tilde{\mu}_{\a_1} \tilde{\mu}_{\a_2} \dots \tilde{\mu}_{\a_n}  \tilde{\phi}^{(+n)} (\tilde{\mu}, \tilde{\nu})\,,  \\
&&
\tilde{\Phi}_{\a_1 \a_2 \dots \a_n}^{(-)} ({p}) = \tilde{\nu}_{\a_1} \tilde{\nu}_{\a_2} \dots \tilde{\nu}_{\a_n}  \tilde{\phi}^{(-n)} (\tilde{\mu}, \tilde{\nu})\,.
\eea
\end{subequations}
Indeed, from eqs.
 \eqref{1.17} and \eqref{1.18} it follows that 
\be
\tilde{p}^{\b \g} \tilde{\Phi}^{(\pm)} _{\b \g \a (n-2)}=0
\label{1.19}
\ee
and eq.~\eqref{1.2.1} is satisfied. Furthermore, using 
\be 
\tilde{p}^{\a \b} \tilde{\mu}_{\b} =- {\rm i} m \tilde{\mu}^{\a}\,, \quad \tilde{p}^{\a \b} \tilde{\nu}_{\b} = {\rm i} m \tilde{\nu}^{\a}\,, 
\label{1.20}
\ee
we obtain
\be 
\tilde{p}^\b{}_{(\a_1}{}
\tilde{\Phi}^{(\pm)}_{\a_2 \dots \a_n) \b} = \mp {\rm i} m  \tilde{\Phi}^{(\pm)}_{\a (n)}\,, 
\label{1.21}
\ee
and so eq.~\eqref{1.2.2} is also satisfied. 
Therefore, $ \tilde{\Phi}^{(\pm)}_{\a (n)} (p) $ 
describe the irreducible massive representations of the Poincar\'e group 
with helicity $\pm n/2$. Since $ \tilde{\Phi}^{(\pm)}_{\a (n)} (p)$ is invariant
under the  transformation \eqref{2.25}, the wave function
$ \tilde{\phi}^{(\pm)} (\tilde{\mu}, \tilde{\nu})$ must  possess the following homogeneity property
\bea
 \tilde{\phi}^{(\pm n)} ({\rm e}^{{\rm i} {\varphi} }\tilde{\mu}, {\rm e}^{-{\rm i} {\varphi} }
 \tilde{\nu}) 
= {\rm e}^{\mp{\rm i} n{\varphi} }
\tilde{\phi}^{(\pm n)} (\tilde{\mu}, \tilde{\nu})\, .
\eea


\section{Projection operators}


Having described the irreducible tensor fields carrying 
definite helicity we can now construct the projection operators onto these states. 

\subsection{On-shell projectors}

We will start with the simplest case of spin $1/2$.
We have two spinors carrying definite helicity
\be 
\tilde{\Phi}^{(+)}_{\a}=\tilde{\mu}_{\a} \tilde{\phi}^{(+)}\,, \qquad 
\tilde{\Phi}^{(-)}_{\a}=\tilde{\nu}_{\a} \tilde{\phi}^{(-)}\,. 
\label{2.0}
\ee
This means that $\tilde{e}^{(+)}_{\a} =\tilde{\mu}_{\a}$ and  $\tilde{e}^{(-)}_{\a} =\tilde{\nu}_{\a}$ are the polarisation spinors. 
Now we define the following projection operators
\be 
\tilde{\Pi}^{(+)}_{\ \a}{}^{\b} =
{\tilde{\mu}_{\a}\tilde{\nu}^{\b}}
\,, \qquad 
\tilde{\Pi}^{(-)}_{\ \a}{}^{\b} =-
{\tilde{\nu}_{\a}\tilde{\mu}^{\b}}
\,. 
\label{2.1}
\ee
They satisfy the following properties
\begin{subequations} \label{2.2}
\begin{align}
\tilde{\Pi}^{(+)}\tilde{\Pi}^{(+)} =  \tilde{\Pi}^{(+)}~&, \qquad \tilde{\Pi}^{(-)}\tilde{\Pi}^{(-)} =  \tilde{\Pi}^{(-)}~, \\
\tilde{\Pi}^{(+)}\tilde{\Pi}^{(-)}&=\tilde{\Pi}^{(-)} \tilde{\Pi}^{(+)}=0~. 
\end{align}
\end{subequations}

Consider an arbitrary on-shell spinor field  $\tilde{\Phi}_{\a} (p) $. 
Then we obtain
\be
\tilde{\Pi}^{(+)}_{\ \a}{}^{\b} \tilde{\Phi}_{\b} 
= 
\tilde{\mu}_{\a} 
\tilde{\nu}^{\b} 
\tilde{\Phi}_{\b}
\,, \qquad 
\tilde{\Pi}^{(-)}_{\ \a}{}^{\b}\tilde{\Phi}_{\b} = 
-\tilde{\nu}_{\a} 
\tilde{\mu}^\b \tilde{\Phi}_{\b}
\, . 
\label{2.3}
\ee
Comparing with eq.~\eqref{2.0} 
we conclude that $\tilde{\Pi}^{(\pm)}$ are the projection operators onto the states with positive and negative helicity. Using the identities
\be 
 \tilde{\mu}_{\a}{\tilde{\nu}^{\b}
 - \tilde{\mu}^{\b} }\tilde{\nu}_{\a} 
= \d_{\a}{}^{\b}\,, \qquad 
 \tilde{\mu}_{\a} {\tilde{\nu}^{\b}
+  \tilde{\mu}^{\b} }\tilde{\nu}_{\a}
= \frac{{\rm i}}{m} \tilde{p}_{\a}{}^{\b}\,, 
\label{2.4}
\ee
we can also write the projection operators in the form 
\be 
\tilde{\Pi}^{(\pm)}_{\ \a}{}^{\b} = \frac{1}{2} \Big( \d_{\a}{}^{\b} \pm \frac{{\rm i}}{m} \tilde{p}_{\a}{}^{\b}\Big)\,. 
\label{2.5}
\ee
At this stage we will remove the tilde assuming that we have performed the transformation to the $\sSL(2, {\mathbb R})$ picture 
$\tilde{\Pi}^{(\pm)} \to \Pi^{(\pm)}, \ \tilde{p} \to p$. 

Now it is clear how to construct the projectors for an arbitrary integer or half-integer spin:
\begin{subequations}\label{2.6}
\bea
&&
\Pi^{(+n)}_{~\a(n)}{}^{\b(n)} = \Pi^{(+)}_{~(\a_1}{}^{\b_1} \cdots \Pi^{(+)}_{~\a_n)}{}^{\b_n}\,, 
 \\
&&
\Pi^{(-n)}_{~\a(n)}{}^{\b(n)}=\Pi^{(-)}_{~(\a_1}{}^{\b_1} \cdots \Pi^{(-)}_{~\a_n)}{}^{\b_n}\,. 
\eea
\end{subequations}
Given  an arbitrary on-shell field $ \Phi_{\a(n)}(p) $, we  define
\be 
\Phi^{(\pm)}_{~\a(n)}=  \Pi^{(\pm n)}_{~\a(n)}{}^{\b(n)}  \Phi_{\b(n)}\,. 
\label{2.7}
\ee
Then it follows that $\Phi^{(\pm)}_{\a(n)}$ satisfies eqs.~\eqref{1.2.1} and~\eqref{1.2.2} and, hence, it  is an irreducible field. This can be checked explicitly
using the identities 
\be 
p_\a{}^{ \b} \Pi^{(\pm)}_{\ \b}{}^{\g}= \mp {\rm i} m 
\Pi^{(\pm)}_{\ \a}{}^{\g}\,, \qquad \ve^{\a \b} 
\Pi^{(\pm)}_{\ \a}{}^{\g} \Pi^{(\pm)}_{\ \b}{}^{\d} =0~.
\label{2.8}
\ee


\subsection{Off-shell projectors}

Let us take a step further and view the projection operators~\eqref{2.5} and \eqref{2.6} as acting not just on the space of on-shell fields,
but on the space of arbitrary fields, whose momentum does not necessarily satisfy $p^2= -m^2$. 
In this case we have to replace $m$ with $\sqrt{-p^2}$, or in the coordinate representation with $\sqrt{\Box}$.

We introduce off-shell projection operators
\be 
{\bm \P}^{(\pm)}_{\ \a}{}^{\b} =\frac{1}{2} \Big( \d_{\a}{}^{ \b} \pm \frac{1}{\sqrt{\Box}} \pa_{\a}{}^{ \b}\Big)\,. 
\label{2.12}
\ee
and their higher-rank extensions (compare  with \eqref{2.6} in the momentum representation)
\begin{subequations} \label{3.13}
\bea
&&
{\bm \Pi}^{(+n)}_{\ \a(n)}{}^{\b(n)} = {\bm \Pi}^{(+)}_{\ (\a_1}{}^{\b_1} \dots 
{\bm \Pi}^{(+)}_{\ \a_n)}{}^{\b_n}\,, 
 \\
&&
{\bm \Pi}^{(-n)}_{\ \a(n)}{}^{\b(n)} = {\bm \Pi}^{(-)}_{\ (\a_1}{}^{\b_1} \dots {\bm \Pi}^{(-)}_{\ \a_n)}{}^{\b_n}\,. 
\eea
\end{subequations}
Given an off-shell field $h_{\a(n)}$, the action of ${\bm \Pi}^{(\pm n) } $ on $h_{\a(n)}$ 
is defined by 
\begin{subequations} \label{3.14}
\bea
{\bm \Pi}^{(+n)} h_{\a(n)} &:=& {\bm \P}^{(+)}_{\ \a_1}{}^{\b_1} \dots {\bm \P}^{(+)}_{\ \a_n}{}^{\b_n}
h_{\b_1 \dots \b_n}  \equiv h^{(+)}_{\a(n)}~, \\
{\bm \Pi}^{(-n)} h_{\a(n)} &:=& {\bm \P}^{(-)}_{\ \a_1}{}^{\b_1} \dots {\bm \P}^{(-)}_{\ \a_n}{}^{\b_n}
h_{\b_1 \dots \b_n}  \equiv h^{(-)}_{\a(n)}~.
\eea
\end{subequations}
The operators ${\bm \Pi}^{(+ n) } $  and ${\bm \Pi}^{(-n) }$ are orthogonal projectors, since 
\bea
{\bm \Pi}^{(+n) } {\bm \Pi}^{(+n) } = {\bm \Pi}^{(+n) } ~, \qquad 
{\bm \Pi}^{(-n) } {\bm \Pi}^{(-n) } ={\bm \Pi}^{(-n) } ~, \qquad 
{\bm \Pi}^{(+n) } {\bm \Pi}^{(-n) } =0~.
\eea

One may also check that the following relations
\bea
\pa^{\a_1 \a_2} {\bm \P}^{(\pm)}_{\ \a_1}{}^{\b_1} {\bm \P}^{(\pm)}_{\ \a_2}{}^{\b_2} =0~,
\qquad
{\bm \P}^{(\pm)}_{~ \a_1}{}^{\b_1} {\bm \P}^{(\pm)}_{\ \a_2}{}^{\b_2} \pa_{\b_1 \b_2}=0
\label{3.16}
\eea
hold. The first identity in \eqref{3.16}  implies that the field $ h^{(\pm)}_{\a(n)}$ is transverse, 
\begin{align}
\partial^{\b \g}h^{(\pm)}_{\b \g \a(n-2)}=0~. \label{2.13.5}
\end{align}
The second identity in \eqref{3.16}  implies that $h^{(\pm )}_{\a(n)}$ is invariant under 
the gauge transformations 
\bea
\d_\z h_{\a(n)} = \pa_{(\a_1\a_2} \z_{\a_3 \dots \a_n)}~. \label{2.13.6}
\eea
In addition to these, one may show that $h^{(\pm )}_{\a(n)}$ satisfies the identity
\begin{align}\label{2.12.5}
\partial^{\b}{}_{(\a_1}h^{(\pm )}_{\a_2\dots\a_n)\b}=\pm\sqrt{\Box}h^{(\pm )}_{\a(n)}~.
\end{align}
The operators ${\bm \Pi}^{(\pm n)}_{\ \a(n)}{}^{\b(n)}$ 
contain terms involving the operator  ${\Box}^{-1/2}$  which requires a special definition.
 However, the sum 
\bea
{\bm \Pi}^{[ n]~~\b(n)}_{\ \a(n)} := {\bm \Pi}^{(+n)}_{\ \a(n)}{}^{\b(n)}  + {\bm \Pi}^{(-n)}_{\ \a(n)}{}^{\b(n)} 
\label{2.13}
\eea
is well defined since it contains only inverse powers of $\Box$ and 
all terms involving  odd powers of ${\Box}^{-1/2}$  cancel out. 
An important observation is that the map
$h_{\a (n)} \to {\bm \Pi}^{[ n ]} h_{\a (n)}$ projects the space of  
symmetric fields $h_{\a(n)}$
onto the space of divergence-free fields, in accordance with \eqref{2.13.5}. 
Thus our projectors \eqref{2.13} are 
the 3D analogues of the Behrends-Fronsdal  projection operators \cite{BF,Fronsdal}.

Furthermore, given an arbitrary field $h_{\a(n)}$, it may be shown that 
\bea
\Big( {\mathbbm 1} - {\bm \Pi}^{[ n]} \Big) h_{\a(n) }
=\pa_{(\a_1 \a_2} \l_{\a_3\dots \a_n)} ~,
\eea
for some $\l_{\a(n-2)}$.

Let $\F_{\a(n)}$ be a field satisfying the Klein-Gordon equation \eqref{mass-shell},
with $n>1$. 
As a consequence of the above analysis, 
the following results hold: 
\begin{enumerate}[(i)]
\item ${\bm \Pi}^{[n] }\F_{\a(n)} $ is a solution of 
the 3D Fierz-Pauli field equations \eqref{1.1.1} and 
\eqref{mass-shell}; and  
\item 
${\bm \Pi}^{(\pm n) } \F_{\a(n)}$ is a solution 
of the equations \eqref{1.1.1} and \eqref{1.1.2}.
\end{enumerate}

We now give several examples of the spin projectors \eqref{2.13}:
 \begin{subequations} \label{3.21}
 \begin{align}
  {\bm \Pi}^{[ 2]} h_{\a(2) } &= \frac{1}{2}\frac{1}{\Box}\Big(\partial_{\a_1}{}^{\b_1}\pa_{\a_2}{}^{\b_2}h_{\b(2)}+\Box h_{\a(2)}\Big)~,\\
  {\bm \Pi}^{[ 3]} h_{\a(3) } &= \frac{1}{2^2}\frac{1}{\Box}\Big(3\partial_{(\a_1}{}^{\b_1}\pa_{\a_2}{}^{\b_2}h_{\a_3)\b(2)}+\Box h_{\a(3)}\Big)~, \\
  {\bm \Pi}^{[ 4]} h_{\a(4) } &= \frac{1}{2^3}\frac{1}{\Box^2}\Big(\pa_{\a_1}{}^{\b_1}\cdots\pa_{\a_4}{}^{\b_4}h_{\b(4)}+6\Box\pa_{(\a_1}{}^{\b_1}\pa_{\a_2}{}^{\b_2}h_{\a_3\a_4)\b(2)}
  +\Box^2 h_{\a(4)}\Big)~,\\
  {\bm \Pi}^{[ 5]} h_{\a(5) } &= \frac{1}{2^4}\frac{1}{\Box^2}\Big(5\pa_{(\a_1}{}^{\b_1}\cdots\pa_{\a_4}{}^{\b_4}h_{\a_5)\b(4)}+10\Box\pa_{(\a_1}{}^{\b_1}\pa_{\a_2}{}^{\b_2}h_{\a_3\a_4\a_5)\b(2)}~~~~~~~~~~~~~~~ \notag \\
 & \phantom{blank~~~~~~}+\Box^2h_{\a(5)}\Big)~,\\
  {\bm \Pi}^{[ 6]} h_{\a(6) } &= \frac{1}{2^5}\frac{1}{\Box^3}
  \Big(\pa_{\a_1}{}^{\b_1}\cdots\pa_{\a_6}{}^{\b_6}h_{\b(6)}+15\Box\pa_{(\a_1}{}^{\b_1}\cdots \pa_{\a_4}{}^{\b_4}h_{\a_5\a_6)\b(4)} \notag \\
& \phantom{blank~~~~~~}+15\Box^2\pa_{(\a_1}{}^{\b_1}\pa_{\a_2}{}^{\b_2}h_{\a_3\dots\a_6)\b(2)}+\Box^3h_{\a(6)}  \Big)~. 
 \end{align}
 \end{subequations} 
 
 All projectors may be rewritten in  vector notation via the standard procedure. Namely, given a bosonic symmetric rank-$(2s)$ spinor field $h_{\a(2s) }$, with integer $s>0$, 
we associate with it the  symmetric rank-$s$ tensor $h_{a_1 \dots a_s}$  
\begin{subequations}
\bea
h_{a_1 \dots a_s}:= \left(-\hf \right)^s(\g_{a_1})^{\a_1 \b_1 } \dots  (\g_{a_s})^{\a_s \b_s } 
h_{\a_1 \b_1 \dots \a_s \b_s} ~,
\eea
which is automatically traceless, 
\bea
\eta^{bc} h_{b c a_1 \dots a_{s-2} } =0~.
\eea 
\end{subequations}
Given a fermionic symmetric rank-$(2s+1)$ spinor field $h_{\a(2s+1) }$, with integer $s>0$, 
we associate with it the  symmetric rank-$s$ tensor-spinor 
$h_{a_1 \dots a_s \g}$ defined by 
\begin{subequations}
\bea
h_{a_1 \dots a_s \g}:= \left(-\hf \right)^s(\g_{a_1})^{\a_1 \b_1 } \dots  (\g_{a_s})^{\a_s \b_s } 
h_{\a_1 \b_1 \dots \a_s \b_s \g} ~.
\eea
It is automatically traceless and $\g$-traceless,
\bea
\eta^{bc} h_{b c a_1 \dots a_{s-2} \g} =0~, 
\qquad (\g^b)^{\b\g} h_{b a_1 \dots a_{s-1} \g } =0~. 
\label{3.21b}
\eea 
\end{subequations}
In vector notation, the examples \eqref{3.21} are equivalent to 
\begin{subequations}
\begin{align}
{\bm \Pi}^{[ 2]} h_{a\phantom{b\a} }&=\frac{1}{\Box}\Big(\Box h_a-\pa_a\pa^b h_b \Big)~,\\
{\bm \Pi}^{[ 3]} h_{a \g \phantom{b} }&=\frac{1}{\Box}\Big( \Box h_{a\g}
-\pa_a\pa^bh_{b\g}-\frac{1}{2}\ve_{abc}(\g^b)_{\g}{}^{\d}\pa^c\pa^dh_{d\d}\Big)~,\\
{\bm \Pi}^{[ 4]} h_{ab\phantom{\a}}&=\frac{1}{\Box^2}\Big(\Box^2 h_{ab}-2\Box\pa^c\pa_{(a}h_{b)c}
+\frac{1}{2}\Box\eta_{ab}\pa^c\pa^dh_{cd}+\frac{1}{2}\pa_a\pa_b\pa^c\pa^dh_{cd}\Big)~,\\
{\bm \Pi}^{[5]} h_{ab\g}&=\frac{1}{\Box^2}\Big(\Box^2h_{ab\g}-2\Box\pa^c\pa_{(a}h_{b)c\g}+\frac{1}{4}\Box\eta_{ab}\pa^c\pa^dh_{cd\g}+\frac{3}{4}\pa_a\pa_b\pa^c\pa^dh_{cd\g} \phantom{~~~~~~~~~~~~~~~~~~~~~~~~~~~}\notag\\
&\phantom{blank sp}-\frac{1}{2}(\g^c)_{\g}{}^{\d}\ve_{cd(a}\big[\Box\pa^d\pa^fh_{b)f\d}-\pa_{b)}\pa^d\pa^f\pa^gh_{fg\d}\big]\Big)~,\\
{\bm \Pi}^{[6]} h_{abc}&=\frac{1}{\Box^3}\Big(\Box^3h_{abc}-3\Box^2 \pa^d\pa_{(a}h_{bc)d}+\frac{3}{4}\Box^2\pa^d\pa^f\eta_{(ab}h_{c)df}+\frac{9}{4}\Box \pa^d\pa^f \pa_{(a}\pa_bh_{c)df}\notag \\
&\phantom{blank sp}-\frac{3}{4}\Box\eta_{(ab}\pa_{c)}\pa^d\pa^f\pa^gh_{dfg}-\frac{1}{4}\pa_a\pa_b\pa_c\pa^d\pa^f\pa^g h_{dfg}\Big)~.
\end{align}
\end{subequations}
One may check that the conditions \eqref{3.21b}  hold.


\section{Linearised higher-spin Cotton tensors} 

Associated with a conformal gauge field $h_{\a(n)}$, with $n>1$, 
is the  linearised Cotton tensor $C_{\a(n)}(h)$  given by the expression 
\eqref{0.3}. 
Its fundamental properties are described by the relations \eqref{0.4} and \eqref{0.5}.
 In this section we derive a new representation for the higher-spin Cotton tensor 
 $C_{\a(n)}$ which makes 
obvious the properties \eqref{0.4} and \eqref{0.5}.

Making use of the spin projection operators, eqs. \eqref{3.13} and   \eqref{3.14}, 
it is possible to show that the following relation holds
 \bea
h^{(\pm)}_{\a(n)} \equiv {\bm \Pi}^{(\pm n)} h_{\a(n)}&=&\frac{1}{2^n}\sum_{j=0}^{n}\binom{n}{j}
\non \\
&& \times
\frac{(\pm 1)^j}{\Box^{j/2}}\partial_{(\a_1}{}^{\b_1}\dots\partial_{\a_j}{}^{\b_j}
h_{\a_{j+1}\dots\a_n)\b_1\dots\b_j}~. \label{4.2}
 \eea
To construct the higher-spin Cotton tensor using the projectors, it is necessary to consider separately the cases of integer and half-integer spin. 
 
 We will begin with the fermionic case and set $n=2s+1$ for integer $s>0$. If we take the sum of the positive and negative helicity parts  of $h_{\a(2s+1)}$, then all terms with odd $j$ in \eqref{4.2} will vanish,
 \bea
 h^{(+)}_{\a(2s+1)}+h^{(-)}_{\a(2s+1)}&=&\frac{1}{2^{2s}}\sum_{j=0}^{s}\binom{2s+1}{2j+1}
 \non \\
 && \times
 \frac{1}{\Box^{s-j}}\partial_{(\a_1}{}^{\b_1}\dots\partial_{\a_{2s-2j}}{}^{\b_{2s-2j}}h_{\a_{2s-2j+1}\dots\a_{2s+1})\b_1\dots\b_{2s-2j}}~.
 \eea
 From here it follows that the fermionic Cotton tensor may be written as
 \begin{align}
 C_{\a(2s+1)}(h)=\Box^s
 \Big({\bm \Pi}^{(+2s+1)} +{\bm \Pi}^{(- 2s-1)} \Big) 
  h_{\a(2s+1)}
 ~.\label{4.4}
 \end{align}

 In the  $n=2s$ case, we instead take the difference of the positive and negative helicity modes, whereupon all even terms in \eqref{4.2} cancel and we obtain
 \bea
  h^{(+)}_{\a(2s)}-h^{(-)}_{\a(2s)}&=&\frac{1}{2^{2s-1}}\sum_{j=0}^{s-1}\binom{2s}{2j+1}
  \non \\
  && \times
  \frac{1}
  {\Box^{(2s-2j-1)/2}}\partial_{(\a_1}{}^{\b_1}\dots\partial_{\a_{2s-2j-1}}{}^{\b_{2s-2j-1}}h_{\a_{2s-2j}\dots\a_{2s})\b_1\dots\b_{2s-2j-1}}~.
 \eea
 Therefore, we may express the bosonic Cotton tensor as
 \begin{align}
C_{\a(2s)}(h)= \Box^{s-\frac{1}{2}}
\Big({\bm \Pi}^{(+2s)} -{\bm \Pi}^{(- 2s)} \Big) 
h_{\a(2s)}~. \label{4.6}
 \end{align}
 By virtue of the identites \eqref{2.13.5} and \eqref{2.13.6}, the properties \eqref{0.4} and \eqref{0.5} are made manifest when $C_{\a(n)}$ is represented in the form \eqref{4.4} and \eqref{4.6}. 
 
Using the identity \eqref{2.12.5}, it is possible to show that the following relations between the derivative of the Cotton tensors and the projectors hold,
\begin{subequations} \label{4.66}
\begin{align}
\pa^\b{}_{(\a_1} C_{\a_2\dots\a_{2s})\b}&=\Box^s\Big({\bm \Pi}^{(+2s)} +{\bm \Pi}^{(- 2s)} \Big) h_{\a(2s)}~,\\
\pa^\b{}_{(\a_1} C_{\a_2\dots\a_{2s+1})\b}&=\Box^{s+\frac 12}\Big({\bm \Pi}^{(+2s+1)} -{\bm \Pi}^{(- 2s-1)} \Big) h_{\a(2s+1)}~.
\end{align}
\end{subequations}
 
 Finally, making use of the relations \eqref{4.4}, \eqref{4.6} and  \eqref{4.66}, in conjunction with the identities
 \bea
 \pa_\a{}^\g {\bm \P}^{(\pm)}_{\ \g}{}^{\b}  = \pm \sqrt{\Box} {\bm \P}^{(\pm)}_{\ \a}{}^{\b} ~,
 \qquad {\bm \P}^{(\pm)}_{\ \a}{}^{\g} \pa_\g{}^\b   
 = \pm \sqrt{\Box} {\bm \P}^{(\pm)}_{\ \a}{}^{\b} ~,
 \eea
 we arrive at the following property 
 \bea
 C_{\a(n)} (\pa h) = \pa_{(\a_1}{}^{\b}C_{\a_2\dots\a_{n})\b} (h)~, \qquad
(\pa h)_{\a(n)} :=  \pa_{(\a_1}{}^{\b}h_{\a_2\dots\a_{n})\b}~.
 \eea


\section{Concluding comments}

In four dimensions, the linearised conformal higher-spin actions 
\cite{FT} were originally formulated in terms of the Behrends-Fronsdal  projection operators \cite{BF,Fronsdal}, and several years later in terms of the linearised higher-spin Weyl tensors
 \cite{FL1,FL2}. In three dimensions, making use of the relations  \eqref{4.4} and \eqref{4.6}
 allows us to rewrite  the linearised conformal higher-spin actions \cite{PopeTownsend,K16}
  \bea
 S^{(n)}_{\rm CS} [h] \propto \ri^n
  \int \rd^3 x\, h^{\a(n)} C_{\a(n)} (h)
  \label{5.1}
\eea
in terms of the spin projection operators.\footnote{The choices  $n=2$ 
and $n=4$ in \eqref{5.1} correspond to a $\sU(1)$ Chern-Simons term
 \cite{Siegel,Schonfeld,DJT1,DJT2} 
and a  Lorentz Chern-Simons term  \cite{DJT1,DJT2}, respectively.} 
 Moreover, making use of \eqref{4.66} also allows us to rewrite the massive higher-spin 
 gauge models\footnote{The bosonic case, $n=2,4,\dots,$  was first described in \cite{BKRTY}.}
 of \cite{BKRTY,KP}
 \bea
S_{\rm{massive}}^{(n)} [h] 
\propto \ri^n
\int \rd^3 x \, {C}^{\a(n) } ( { h})
\Big\{ \pa^\b{}_{\a_1} - m \s \d^\b{}_{\a_1}  \Big\}  { h}_{\a_2 \dots \a_n \b}~,
\qquad  \s =\pm 1
\label{5.2}
\eea
in terms of the spin projection operators. The Bianchi identity \eqref{0.5} 
and the equation of motion derived from \eqref{5.2} are equivalent 
to the massive equations \eqref{1.1.12}.

In the $n=2$ case, the action \eqref{5.2} proves to be proportional to that for topologically massive electrodynamics 
 \cite{Siegel,Schonfeld,DJT1,DJT2} 
\bea
S = -\frac 14 \int \rd^3 x \, \Big\{ F^{ab}F_{ab} +m \s \ve^{abc}h_a F_{bc}\Big\}~,
\qquad F_{ab} = \pa_a h_b - \pa_b h_a~.
\eea
One may check that for $n=4$, the action \eqref{5.2} yields 
 linearised new topologically massive gravity 
\cite{ABdeRHST,DM}.

It should be pointed out that various aspects of the bosonic higher-spin 
Cotton tensors $C_{\a(n)}$, with $n $ even, were studied in \cite{LN,BBB}.

The results of this work admit supersymmetric extensions. 
They  will be discussed elsewhere. 
\\


\noindent
{\bf Acknowledgements:} SMK is grateful to Alexey Isaev and Arkady Segal 
for useful discussions.
This work is supported in part by the Australian 
Research Council, project No. DP160103633.
The work of MP is supported by the Hackett Postgraduate Scholarship UWA,
under the Australian Government Research Training Program.


\appendix


\section{Spinor conventions}


Here we summarise our  notation and spinor conventions which follow \cite{KLT-M11}.
 We use the metric $\eta_{mn}= {\rm diag} (-1, 1, 1)$ and normalise the Levi-Civita symbol as $\ve^{012}=-\ve_{012}=1$. 

In the $\sSL(2, {\mathbb R})$ picture, the $\g$-matrices with lower indices 
 are chosen as
\be 
(\g_m)_{\a \b} = (\g_m)_{\b \a} = ({\mathbbm 1}, \s_1, \s_3)\,. 
\label{A1}
\ee
The spinor indices are raised and lowered, 
\bea
\psi^{\a}=\ve^{\a\b}\psi_\b~, \qquad \psi_{\a}=\ve_{\a\b}\psi^\b~,
\label{A.22}
\eea
 using the antisymmetric tensors 
$\ve_{\a \b}= -\ve_{\b\a} $ and $\ve^{\a \b} =- \ve^{\b\a} $ normalised as $\ve_{12}=-1$
and $ \  \ve^{12}=1$. 
The Dirac $\g$-matrices are
\be 
(\g_m)_{\a}{}^{\b}:= \ve^{\b \g} (\g_m)_{\a \g}  = (-{\rm i} \s_2, \s_3, \s_1)\,.
\label{A2}
\ee
The $\g$-matrices have  the following properties: 
\begin{subequations}\label{A3}
\bea
(\g_m)_{\a}{}^{ \rho} (\g_n)_{\rho}{}^{ \b}  
&=& \eta_{mn} \d_{\a}{}^{ \b}  +\ve_{mnp} (\g^p)_{\a}{}^\b
\,, 
 \\
(\g^m)_{\a \b} (\g_m)^{\rho \s}  &=& - (\d_{\a}{}^{ \rho} \d_{\b}{}^{ \s}  + \d_{\a}{}^{ \s} \d_{\b}{}^{ \rho})\,, 
\\
\ve_{amn}(\g^m)_{\a\b}(\g^n)_{\g\d}&=&
\ve_{\g(\a}(\g_a)_{\b)\d}
+\ve_{\d(\a}(\g_a)_{\b)\g} \, .
\eea
\end{subequations}

Given a three-vector $\F_a$, it can equivalently be realised as a symmetric rank-2  spinor 
$\F_{\a\b} =\F_{\b \a}$.
The relationship between $\F_a$ and $\F_{\a \b}$ is as follows:
\bea
\F_{\a\b}:=(\g^a)_{\a\b}\F_a~,\qquad
\F_a=-\hf(\g_a)^{\a\b}\F_{\a\b}~.
\label{vector-rule}
\eea

In the $\sSU(1, 1)$ picture, the $\g$-matrices with lower indices are  
\be 
(\tilde{\g}_m)_{\a \b}= (T^{-1})_{\a}^{\ \g}(T^{-1})_{\b}^{\ \d} (\g_m)_{\g \d}\,, 
\label{A5}
\ee
where $T$ is given by~\eqref{1.4}. The explicit expressions for these matrices are
\be 
(\tilde{\g}_m)_{\a \b} = (-{\rm i} \s_1, -{\rm i} {\mathbbm  1}, \s_3)\,. 
\label{A6}
\ee
For the Dirac $\g$-matrices we obtain
\be 
(\tilde{\g}_m)_{\a}{}^{ \b}= \ve^{\b \g} (\tilde{\g}_m)_{\a \g}  = (-{\rm i} \s_3, -\s_2, -\s_1)\,.
\label{A7}
\ee
%


\begin{footnotesize}

\end{footnotesize}

\end{document}